# Nanophotonic structures with optical surface modes for tunable spin current generation

P.V. Shilina[1,2*], D.O. Ignatyeva[3,4], P.O. Kapralov[3], S.K. Sekatskii[5], M. Nur-E-Alam[6], M. Vasiliev[6], K. Alameh[6], V. G. Achanta[7], Y.Song[8], S.M. Hamidi[9], A.K. Zvezdin[1,2] and V.I. Belotelov[3,4]

[1]*National Research University Higher School of Economics, Moscow 101000, Russia*
[2]*Moscow Institute of Physics and Technology, Dolgoprudny, 141700, Russia*
[3]*Russian Quantum Center, Skolkovo, Moscow 121205, Russia*
[4]*Faculty of Physics, Lomonosov Moscow State University, Moscow 119991, Russia*
[5]*Institute of the Physics of Biological Systems, École Polytechnique Fédérale de Lausanne, Lausanne CH-1015, Switzerland*
[6]*Electron Science Research Institute, Edith Cowan University, Joondalup 6027, Australia*
[7]*Tata Institute of Fundamental Research, Mumbai 400005, India*
[8]*University of Science and Technology Beijing, Beijing 100083, China*
[9]*Laser and Plasma Research Institute, Shahid Beheshti University, Evin Tehran 19839 69411, Iran*

\* *polly.penkina@gmail.com*

Heat generated by spin currents in spintronics-based devices is typically much less than that generated by charge current flows in conventional electronic devices. However, the conventional approaches for excitation of spin currents based on spin-pumping and spin Hall effect are limited in efficiency which restricts their application for viable spintronic devices. We propose a novel type of photonic-crystal (PC) based structures for efficient and tunable optically-induced spin current generation via the Spin Seebeck and inverse spin Hall effects. It is experimentally demonstrated that optical surface modes localized at the PC surface covered by ferromagnetic layer and materials with giant spin-orbit coupling (SOC) notably increase the efficiency of the optically-induced spin current generation and provides its tunability by modifying light wavelength or angle of incidence. Up to 100% of the incident light power can be transferred to heat within the SOC layer and, therefore, to spin current. Importantly, high efficiency becomes accessible even for ultra-thin SOC layers. Moreover, surface patterning of the PC-based spintronic nanostructure allows local generation of spin currents at the pattern scales rather than diameter of the laser beam.

<u>Keywords</u>: photonic-crystal, spin Seebeck effect, inverse spin Hall effect, spin current.

## I. INTRODUCTION

Nowadays electronics has reached its fabrication limits using conventional semiconductor foundries, and there is a need to use alternative technologies for further improve the performance of electronic devices in terms of size, power consumption, heat dissipation and stability. Spintronics is a modern and rapidly developing field and currently considered one of the most promising for data storage and processing [1,2]. Pure spin currents allow for transferring spins without charge transfer, and this significantly reduces the losses associated with Joule heating and, consequently, the power consumption [3,4]. Current research on spintronics is focused on the problems of spin current generation, detection and the viable applications of spintronics.

Various mechanisms of spin-current generation have been proposed, such as spin-pumping based on ferromagnetic resonance (FMR) [5,6], and Spin Hall effect whereby spin currents perpendicular to charge currents appear in materials with a strong spin-orbit coupling (SOC) [7-9]. Recently, Optical techniques have been proposed for the efficient generation spin-current, and even the term of optospintronics has commonly been used [10]. Optospintronics is based on the use of femtosecond laser pulses to excite spin currents. Circular polarized laser pulses can launch spin currents in a semiconductor non-thermally [11]. Another optospintronics approach is based on the quantum interference of one- and two-photon absorption of femtosecond phase-locked pulses having orthogonal linear polarizations [12]. Furthermore, optical spin-pumping has been used for spin-current generation [13].

One of key optospintronics features for pure spin current generation is the thermal optical mechanism [14-16]. Illumination of the sample with a laser beam creates a temperature gradient in it, which results in spin current flow along the temperature gradient in the presence of a magnetic field, due to the Spin Seebeck effect (SSE) [17-20]. Although thermal mechanism is slower than spin-pumping, it is important that it does not require complex and high-power equipment like femtosecond laser, and thus is more promising for usage in commercial spintronic devices where device size and power consumption play a crucial role.

For the efficient generation of the spin current by laser heating via SSE, it is necessary to absorb most of the incident light in order to heat the magnetic layer strongly and non-uniformly, thus establishing a large thermal gradient. However, in the smooth magnetic films studied so far [14,15], there is not much room for manipulation with the thermal gradient. Indeed, for transparent magnets, the light is distributed mostly uniformly, whereas for metallic magnets, the light decays simply exponentially. Moreover, a relatively large part of the incident optical power is not absorbed. In this paper, we show that the situation might be improved if light is trapped inside or in the vicinity of the magnetic layer through appropriate excitation of the optical mode.

Here, we demonstrate that the use of nanophotonics is very promising as it enables (i) better concentration of the optical power at a subwavelength scale in the necessary region of the material [21,22] and (ii) tuning of the system, due to the high-Q of the resulting resonant nanostructure [23-25]. In particular, we propose spintronic nanostructures consisting of a photonic crystal (PC), a ferromagnetic layer and a layer with giant SOC, as shown in Fig. 1a. Optical modes arising from the periodicity disturbance at the PC surface are often referred as optical Tamm states [26-28]. A crucial property of a PC is that its surface mode is concentrated near the air interface [29-32], and this is extremely important to form high temperature gradient. Moreover, the long propagation length of surface modes in conjunction with the very high-Q resonance increases the local light intensity by several orders of magnitude [31]. It has been shown that such surface modes in PC-based structures with magnetic materials, such as iron-garnet and cobalt, are very sensitive to the magnetization [27, 23-25]. In some sense, the excited surface waves are similar to the surface plasmon-polaritons [33-35], but they do not require thick metals, and hence, they high-Q and tunable resonances can be attained using few-nm-thick films of spintronic materials such as Ta, Py ($Ni_{80}Fe_{20}$) or Pt, whose optical losses are rather high.

In this work, two different PC-based spintronic nanostructures are considered: one is covered with bismuth-substituted iron-garnet (BIG) layer and Pt nanofilm (the BIG+Pt sample), and the other one is covered with Py ($Ni_{80}Fe_{20}$) and Ta nanofilms (the Py+Ta sample) (Fig. 1b). The BIG+Pt nanostructure exhibits superior performance due to higher temperature gradient induced by light, and hence, it possible to achieve high efficiency thermal optical spin-current generation possible using this nanostructure. Moreover, the extremely thin, less than 2 nm thick, Pt films provides another degree of freedom for device tunability and miniaturization via additional nanopatterning.

## II. OPTICAL MODE EXCITATION IN SPINTRONIC PC-BASED NANOSTRUCTURES

The main concept of this work is to design a proper spintronic-nanophotonic structure that concentrates the incident optical power within an ultrathin magnetic layer, thus establishing temperature gradient high enough to launch spin currents via the spin Seebeck (SSE) or a similar effect. PC-based structures can sustain a surface optical mode due to the phenomenon of total internal reflection from the interface with air on the bottom side, and a bandgap of the PC on the top side.

In our proposed spintronic PC-based nanostructures (BIG+Pt and Py+Ta (Fig. 1b)) the pairs of a magnetic film and a film with larger SOC are introduced to generate spin currents via SSE and detect them via the inverse spin Hall effect. The BIG+Pt sample consists of a PC with 16 pairs of substructures, each consist of four layers, namely, a 119.4 nm thick $Ta_2O_5$, a164.4 nm thick $SiO_2$, a 125 nm thick iron-garnet layer and a 3-nm-thick Pt layer. On the other hand, the Py+Ta sample is a PC with 14 pairs substructures, each comprises a 119.4 nm thick $Ta_2O_5$, a 164.4 nm thick $SiO_2$, a 107.5 nm thick $Ta_2O_5$, a 7-nm-thick Py and a 7-nm-thick Ta layer (see Supplementary S1).

To excite the surface modes, the lateral momentum of the incident light should match the momentum of the surface mode. This is achieved by illuminating the samples through the $SiO_2$ prism attached from the substrate side, using an appropriate index matching oil (Fig. 1a). The sample is illuminated by a laser diode of wavelength 805 nm and output power $W$=50 mW. The intensity of the reflected light is measured by a photodetector for different angles of incidence (from 30 to 60 degrees) and is normalized to the intensity of the incident light. An external magnetic field is applied along the transversal direction. The spin current, which appears due to laser heating by the mechanically modulated light in the presence of a alternating external magnetic field, is detected by measuring the magnitudes of laser-induced voltages at the sum of the external magnetic field frequency and the chopper frequency (see details in Supplementary S2).

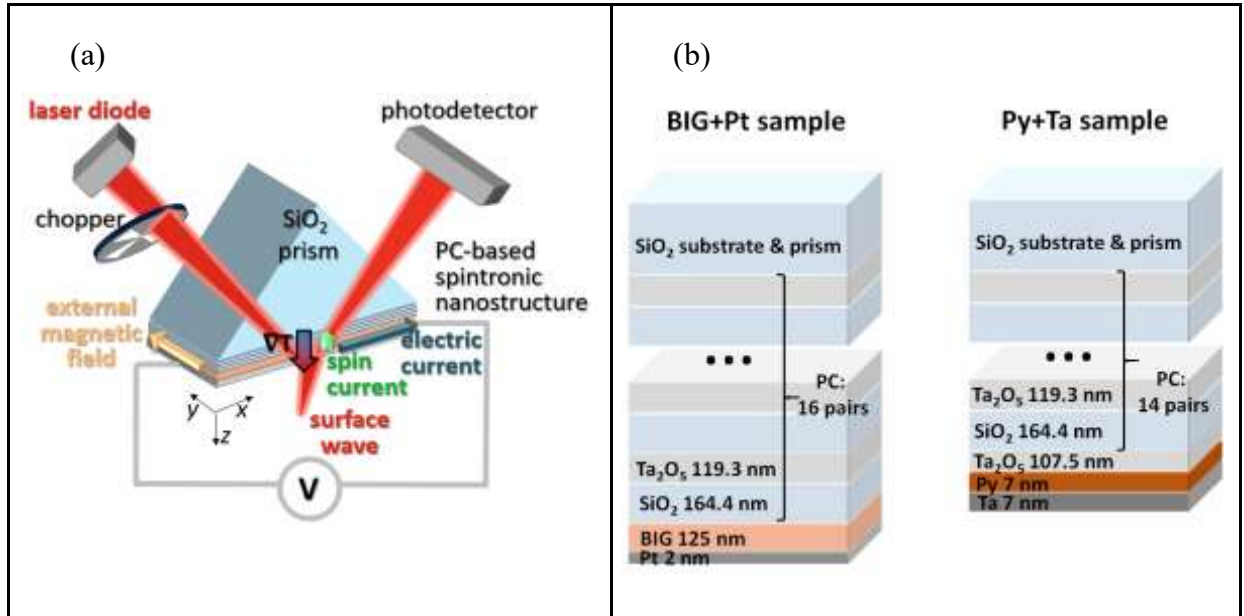

**Figure 1.** Spin current excitation in the PC-based nanostructures with surface optical modes. (a) Schematics of the experimental set-up for reflectance spectra and optically induced spin current measurement; (b) Structures of the BIG+Pt and Py+Ta samples, brown color corresponds to the ferromagnetic material, and dark gray color -to the material with giant spin-orbit coupling.

The parameters of the PC were selected in such a way that the center of the PC bandgap corresponds to an angle of the optical mode excitation close to the total internal reflection angle, $\theta_{TIR}$=43.23 deg, at the operating wavelength of 805 nm (horizontal green dashed line in Fig. 2c,d). Figures 2 (a and b) show reflectance (black line) and absorption (red line) angular spectra obtained experimentally (solid lines) and numerically (dotted lines). Figures (c and d) show density plots of the reflectance versus wavelength and angle of incidence, for the BIG+Pt (left) and the Py+Ta (right) nanostructures. The PC bandgap corresponding to the spectral region with R≈100% is clearly seen for both structures in the angular spectra of reflection and absorption, which were calculated using the impedance method (dotted lines in Figs 2a,b and density plots in Figs 2c,d) described in the Supplementary S3 section. Smaller magnitude peaks of the reflectance coefficient surrounding the bandgap correspond to the Fabry-Perot interference resonances in the PC, governed by the equation $\Sigma d_j n_j / \cos(\theta_j) = (2m + 1)\lambda/2$ (where $d_j$, $n_j$ and $\theta_j$ are the layer thicknesses, refractive indices and the angles inside the corresponding layers of the PC, $m$ is an integer). One should note that in the non-absorbing structures $R$=100% for $\theta > \theta_{TIR}$, however, since the Pt layer is highly absorptive, there are also interference resonances in this angular range. The surface optical mode (blue dashed curve in Figs 2c,d) for $\theta > \theta_{TIR}$, can be excited over a wide angular range, however, the corresponding resonance becomes significantly narrow only around the total internal reflection angle.

The measured spectra are in a good agreement with the numerically simulated ones (Fig. 2a,b). Excitation of the optical surface mode reveals itself as a narrow reflectance dip of width around 0.8 deg for the BIG+Pt sample and 0.1 deg for the Py+Ta sample. Numerical simulations predict that the resonance could be even narrower, however, focusing of the laser diode to the beam with an angular width of 0.5 deg comparable with the resonance angular width made it wider and shallower.

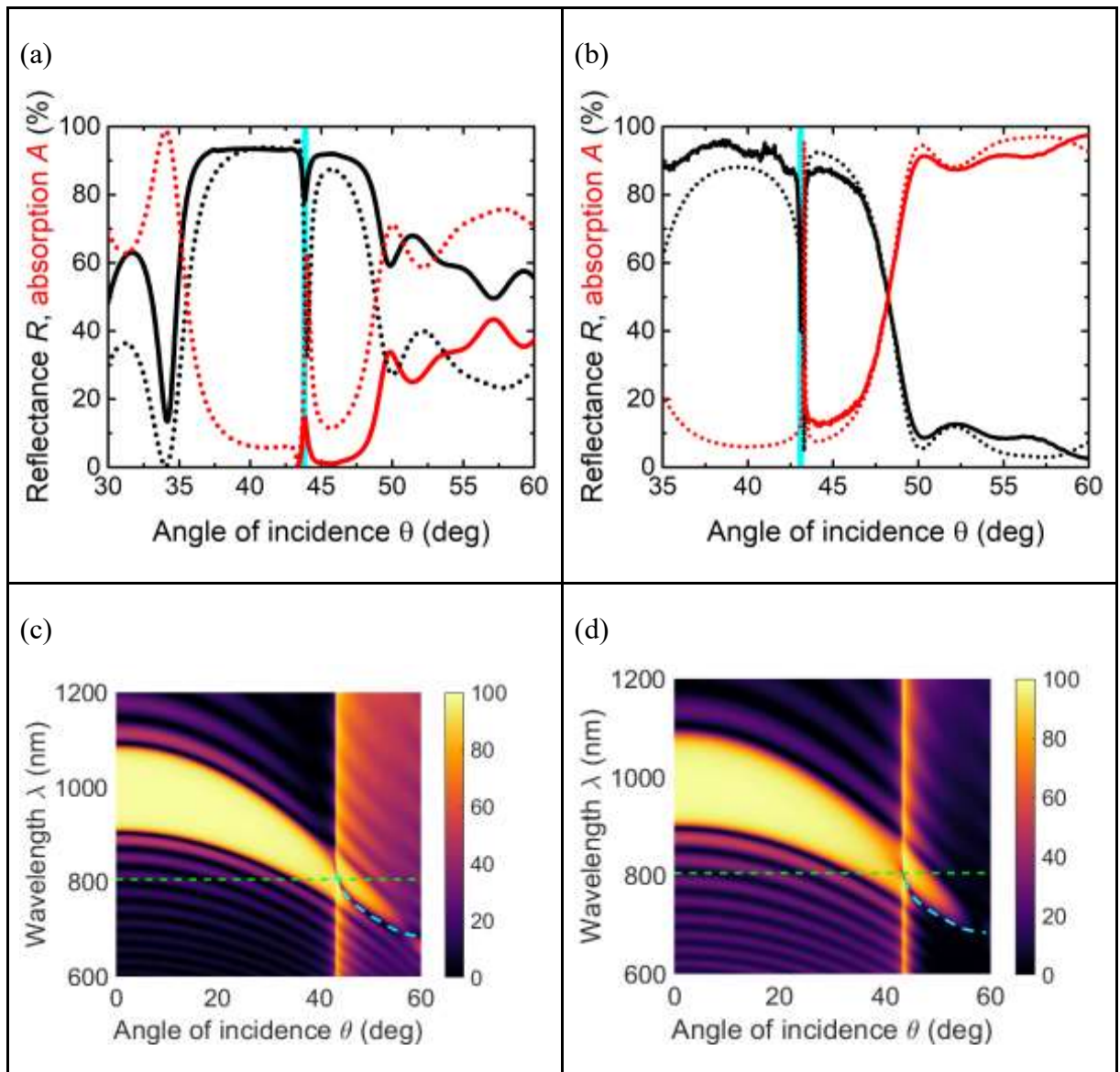

**Figure 2.** Optical properties of the BIG+Pt (left pane) and the Py+Ta (right pane) samples. (a,b) Reflectance (black line) and absorption (red line) angular spectra obtained experimentally (solid lines) and numerically (dotted lines). Experimental spectra of absorption were calculated only for the total internal reflection angular range with the assumption $A=1-R-S$. (c,d) Density plots of the calculated reflectance coefficient versus the wavelength and angle of incidence. Green dashed line shows the operating wavelength $\lambda=805$ nm. Light-blue dashed lines show the surface optical mode contour.

## III. SPIN CURRENTS GENERATED BY THE LASER BEAMS

Since the considered nanostructures demonstrated excellent concentration of the optical energy within the magnetic layer, they can be used for efficient generation of spin currents by light. Indeed, illumination of the sample with a laser light induces temperature gradient along the z-axis (Fig. 1a). Iron-garnet, as well as the PC layers $SiO_2$ and $Ta_2O_5$ are rather transparent materials at 805 nm wavelength ($\varepsilon$"<10-3). Thus, regardless of the light distribution inside the PC-based nanostructure, according to the Joule-Lenz law, light is predominantly absorbed in the lossy metallic films, i.e., in the Pt film for the the BIG+Pt sample and in the Py and Ta films for the Py+Ta sample. Some inhomogeneity of the electromagnetic field distribution inside the metal films was seen, however, it was insignificant since metals typically possess about 50-100 times higher thermal conductivity than $SiO_2$ and $Ta_2O_5$, and about 2500 times higher than air. Thus, the bottom metallic film illuminated by the light plays a key role as a heating source, and the energy flux directed towards the substrate provide large thermal gradient $\nabla T$ in the whole sample directed towards the bottom metal layer.

The temperature gradient in the ferromagnetic layer causes the spin current flowing parallel to $\nabla T$ due to the SSE [20]. Since the ferromagnetic layer is in contact with the layer possessing giant SOC (Pt or Ta), the charge current flows perpendicular to the spin current and also to the spin polarization, due to the inverse spin Hall effect (ISHE) [36-38]. The voltage associated with ISHE is given by

$$V_{ISHE} = -\frac{1}{2}\alpha\eta S_s L \nabla T, \qquad (1)$$

where $S_s$ is the spin Seebeck coefficient, $\eta$ is the coefficient of spin injection, $L$ is the distance between the contacts (along x-axis), $\alpha$ is the angle of the Spin Hall effect and $\nabla T$ is the thermal gradient along z-axis.

Note that another effect, of similar symmetry, arises if the magnetic field and temperature gradient are applied to the ferromagnetic layer, and this effect should be taken into account. The anomalous Nernst effect (ANE) is a thermoelectric phenomenon whereby a transverse electric field is generated perpendicular to both magnetization and applied temperature gradient. The voltage induced by the ANE [39,40] is given by

$$V_{ANE} = -S_{ANE} L \nabla T, \qquad (2)$$

where $S_{ANE}$ is the ANE coefficient. Usually ANE is observed in ferromagnetic metals [41, 42]. Consequently, its contribution is important for Py layer of Py+Ta structure. However, for the BIG+Pt sample, the ANE contribution is almost zero due to the absence of charge carriers [43]. Note that charge carriers have been observed in Pt+BIG structures due to the magnetic proximity effect in Pt being in contact with BIG [44], however, recent experimental studies of such structures have demonstrated that the number of these carriers is relatively insignificant [45].

Therefore, both the ANE and ISHE contribute to the observed optically-induced voltage, i.e., $V=V_{ANE}+V_{ISHE}$ (see Fig. 3).

The induced voltage for both samples becomes saturated if the applied magnetic fields are high enough to saturate the sample magnetization (Fig. 3ef). For the lower magnetic field, the dependence $V(H_{ext})$ is determined by the hysteresis of the samples. The Py+Ta sample has in-plane magnetic anisotropy, while the anisotropy of the BIG+Pt sample is out-of-plane, therefore their hysteresises in the in-plane magnetic field are different. As a result, $V(H_{ext})$ for the Py+Ta is a step-like function while for the BIG+Pt sample the voltage gradually increases with the magnetic field.

Both the ISHE and ANE depend on the thermal gradient formed in the ferromagnetic layer, which is proportional to the heat flux, according to the well-known Fourier's law of heat conduction. Since the heat flux is proportional to the absorbed optical power, one may assume:

$$V = \beta\, W A\,, \tag{3}$$

where $W$ is the incident laser power and $A$ is the absorption coefficient.

Experimental angular spectra of the induced voltage $V(\theta)$ were measured (blue spheres in Fig.3 a,b). A narrow resonant peak observed at almost zero background corresponds to the excitation of the optical surface mode in the PC bandgap. Other regions where the notable voltage is generated correspond to interference-enhanced light absorption, as discussed above.

In order to quantitatively describe this process, one should take into account the angular dependence of the laser light absorption. As long as we focus our attention on the optical mode that could be excited only in the total internal reflection angular range, one can assume that the transmittance $T=0$, thus the reflectance $R$, absorption $A$ and spurious scattering $S$ due to the surface roughness could be calculated using the relationship: $R+A+S=1$. Both scattering and absorption could not be measured in the present setup, however, their contribution could be estimated from the $R(\theta)$ measurements (Fig.2a,b) As follows: While the absorbed laser power is transformed into heat creating a temperature gradient that lies at the root of the observed ISHE and ANE voltages according to Eqs.(1) and (2), the scattered radiation is not involved in these processes. Then, one can assume that $A(\theta)$ has a resonance features corresponding to the optical modes, while $S(\theta)=S_0$ is a constant value. By linearly fitting the voltage-reflectance dependence (Fig. 3c,d), one can obtain simultaneously the scattering $S_0$ and the constant $\beta$ using this linear relationship:

$$V = \beta \cdot W \cdot (1 - R(\theta)) - \beta \cdot W \cdot S_0 \tag{4}$$

Using Eq. (4), the following values were obtained: for the BIG+Pt sample, $S_0=7\%$ and $\beta=9.5$ nV/mW, while for the Ta+Py sample, $S_0<1\%$, $\beta=6.3$ nV/mW. For the Py+Ta structure the induced voltage is slightly higher than that generated by the BIG+Pt structure (Fig.3 a,b), however, $\beta$ is 1.5 times higher for the BIG+Pt structure. For the Py+Ta sample, the ANE contribution to the induced voltage prevails over the ISHE contribution [41, 46]. Consequently, we can conclude, that the product $\alpha\eta S_s$ describing the spin-current generation process (which is proportional to $\beta$), is significantly higher for the BIG+Pt structure, making it superior and more practical for spintronic applications.

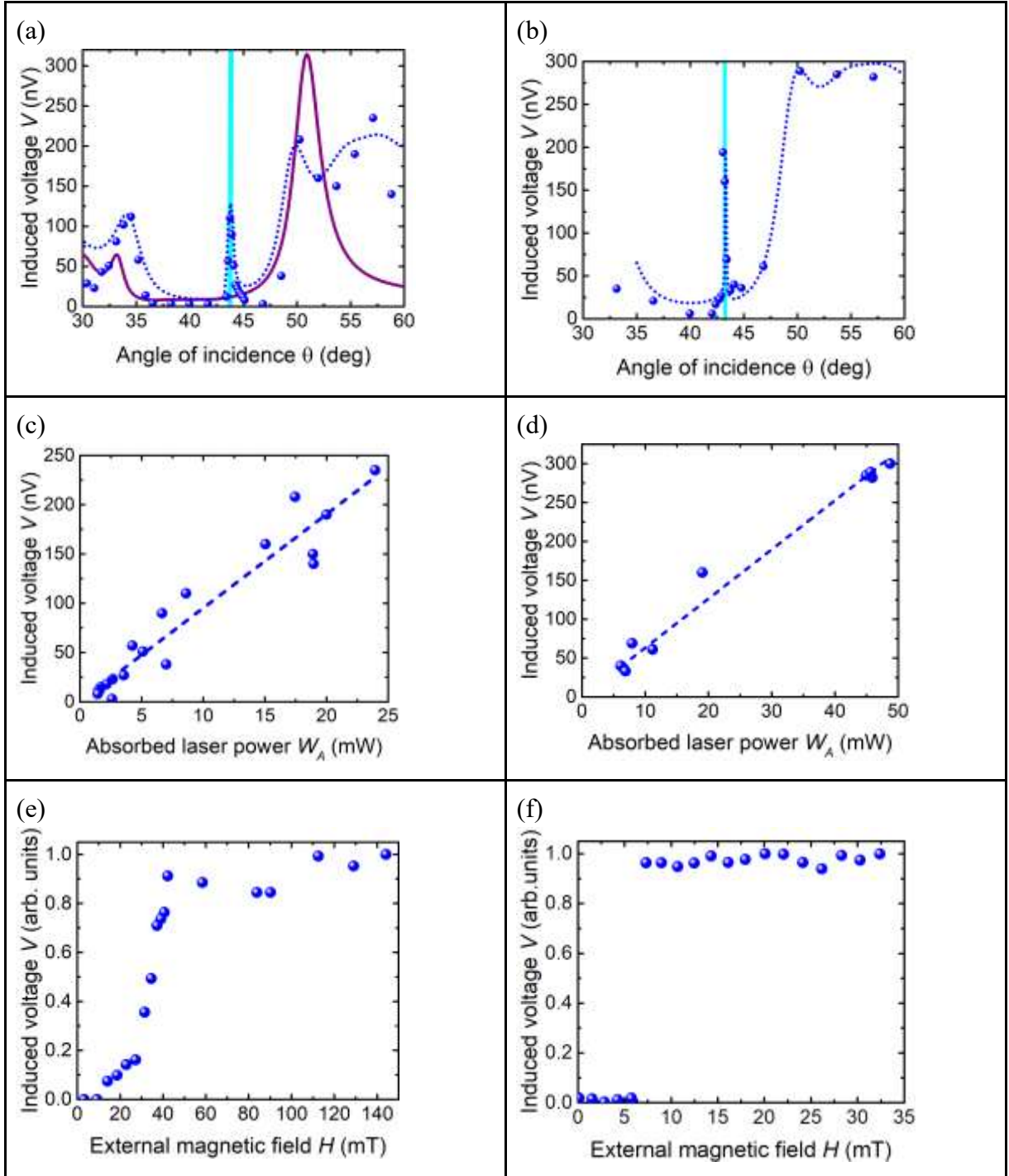

**Figure 3.** Induced voltage arising from optically generated spin currents in BIG+Pt (left pane) and Py+Ta (right pane) samples. (a),(b) Measured (blue dots) and modelled (blue dotted lines) induced voltage angular spectra. Light-blue line show the position of the excited mode. Purple line shows the ISHE for the optimized spintronic BIG+Pt nanostructure. (c),(d) Linear fit of the measured induced voltage dependence on the absorbed laser power. (e),(f) The dependence of the induced voltage on the applied external magnetic field.

In Fig. 3a,b the blue dashed line shows the approximate induced voltage based on the estimated $\beta$ constants and numerically obtained $A(\theta)$ averaged over the corresponding beam angular width. The proposed model is in a good agreement with the experimental results shown by blue dots, and therefore, it can be used for further optimization of the spintronic nanostructure performance.

Note that there is a tade-off between structure tunability provided by the high-Q and the efficiency (limited by the angular width of the laser beams used in experimental setup). Typically, one may tune the spintronic nanostructure parameters to achieve simultaneously wider and deeper resonances with A~100% (see Supplementary S3). The purple curve in Fig.3a shows the ISHE angular spectra for such optimized structure (see Supplementary S3 for parameters) displaying a well-resolved peak associated with the surface optical mode. The resonance features 1 deg angular width and 2.7 nm width in the wavelength domain (Q-factor is 315) and an absorptance of 95%.

## IV. ADVANTAGES OF THE SPINTRONIC PC-BASED NANOSTRUCTURE

Now let us discuss the advantages of the proposed spintronic PC-based nanostructures in more detail. First, as it was mentioned earlier, by tuning the parameters of the PC one may achieve the depth of the resonance caused by the surface optical wave up to 100% in magnitude. This means that up to 100% of the incident light intensity could be transferred to heat within the sample, thus creating the spin currents with the highest possible optical-to-thermal conversion efficiency. Moreover, such high efficiency could be obtained for very thin, e.g., 1-nm-thick, Pt layers. Fig. 3a (purple line) illustrates how the structure efficiency could be maximized when the parameters of the PC-based spintronic structure are optimized (width of BIG layer is 170 nm, width of Pt nanofilm is 1 nm). For comparison, a smooth BIG film covered with 1-nm Pt layer absorbs only 20% of the energy of the incident laser beam, which means that 80% of the laser energy does not contribute to the spin current, and hence, is just wasted.

Another advantage of the developed spintronic nanostructure is that the spectral position of the optical surface mode resonance could be tuned by varying the thickness of the terminating dielectric layer, since the excited optical mode is similar to the guided one [25]. This tunability opens an opportunity for further miniaturization of the optically-operated spintronic devices. For example, the operating wavelength of the diode laser used in the present experiments could be altered in the range of 802-812 nm via the temperature stabilization circuitry. Thus, one may use one diode laser to operate a PC with step like terminating dielectric layer for the excitation of optical surface waves at different steps of the structure by varying wavelengths (Fig. 4a). For example, to achieve the multi-channel spin-current generation at laser wavelengths of 802 nm, 805.5 nm, 808 nm and 811.5 nm one should make a structure with BIG layer steps of heights equal to 117.5 nm, 120 nm, 122.5 nm, and 125 nm, respectively (Fig. 4b).

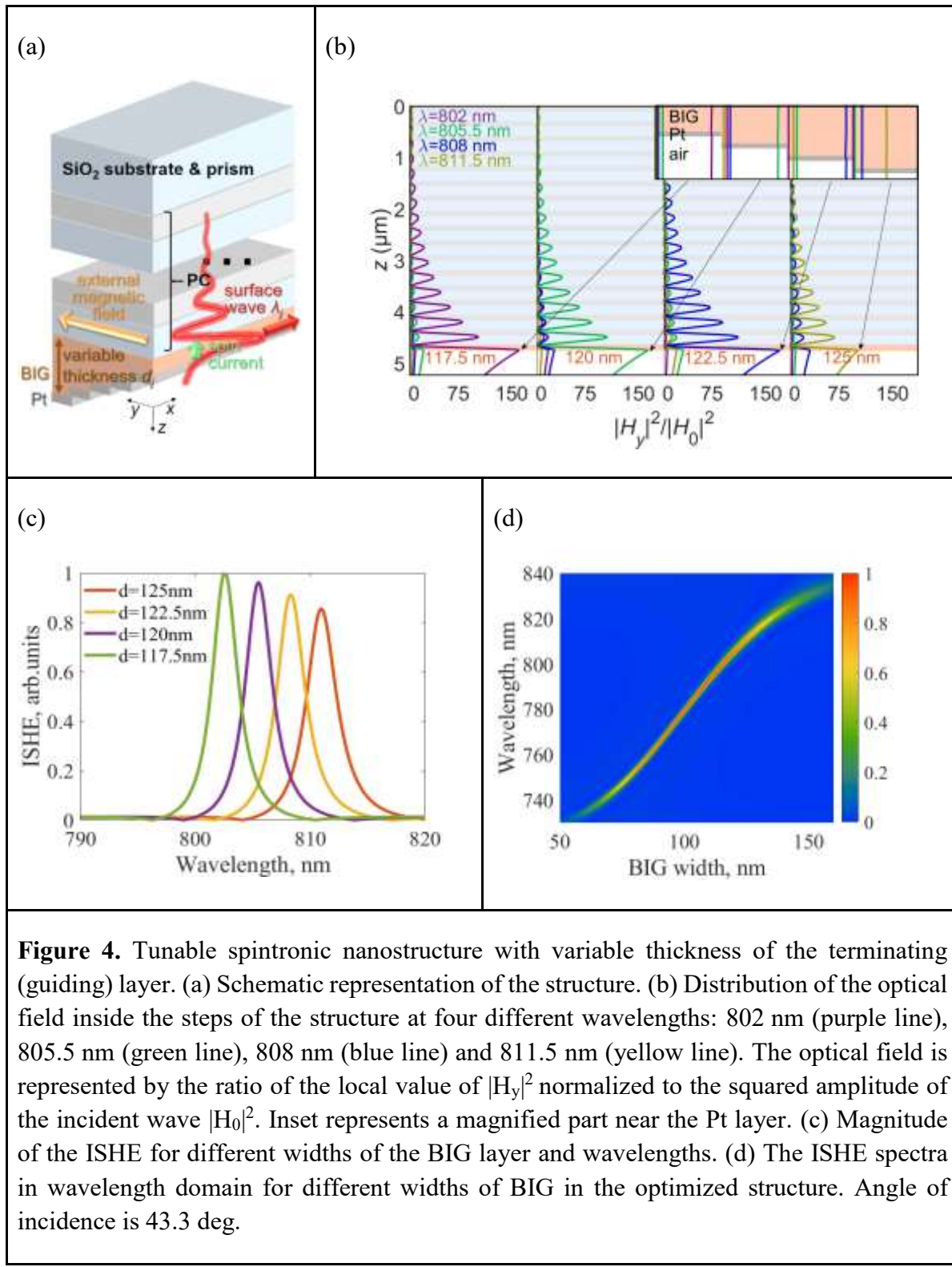

**Figure 4.** Tunable spintronic nanostructure with variable thickness of the terminating (guiding) layer. (a) Schematic representation of the structure. (b) Distribution of the optical field inside the steps of the structure at four different wavelengths: 802 nm (purple line), 805.5 nm (green line), 808 nm (blue line) and 811.5 nm (yellow line). The optical field is represented by the ratio of the local value of $|H_y|^2$ normalized to the squared amplitude of the incident wave $|H_0|^2$. Inset represents a magnified part near the Pt layer. (c) Magnitude of the ISHE for different widths of the BIG layer and wavelengths. (d) The ISHE spectra in wavelength domain for different widths of BIG in the optimized structure. Angle of incidence is 43.3 deg.

At the first step of 117.5 nm height the optical field attains its maximal values in Pt layer at 802 nm (purple line in Fig. 4b) while the field at the other wavelengths is much smaller. Actually, the field enhancement factor is more than 300 with respect to the incident light. Note that the corresponding ISHE resonance peaks are separated just by $\Delta\lambda=2.5$nm, however they do not overlap and the crosstalk from adjacent curves at the resonant wavelength is more than

5 times (7dB) (Fig. 4c). Higher number of operating wavelengths could be achieved via utilization of set of laser diodes with wavelengths separated by around 3 nm over a wide wavelength range (see Fig.4d), e.g., from 750 nm to 820 nm.

Consequently, the area of the spin current generation in the step structure is determined by the widths of the steps rather than diameter of the focused laser beam. It solves the problem of narrow focused laser beams which broaden optical resonances and significantly diminish power concentration in the surface modes. Indeed, using the step like structure allows to deal with very broad beams close to a plane wave and to keep spatial resolution at the micron scale determined by the widths of the steps.

It should be noted that in Fig. 4a the step widths are a bit larger than the laser wavelength. If one considers a structure nanopatterned with stripes (i.e., a diffraction grating) having a spatial scale compared to the light wavelength, it is necessary to take into account the diffraction of light caused by such grating. In this case, it is possible to excite the surface optical mode through the formed diffraction grating, thus eliminating the need for a prism and making the device more compact.

## V. CONCLUSION

We have demonstrated that nanophotonic structures can be used for spintronics. Experimental studies of the specially designed photonic-crystal-based opto-spintronic structures demonstrated that efficiency of the spin current generation can be significantly increased if optical surface modes are excited. The surface modes provide concentration of the optical power within the SOC layer and enhance the optically-induced thermal mechanism for spin-current generation via the Spin Seebeck and inverse spin Hall effects. Up to 100% of the incident light power can be transferred to heat and, therefore, to spin current. Thickness of the SOC layer can be made ultra-small down to 1 nm.

It is possible to control the value of excited spin currents not only by incident light intensity but also by its wavelength and angle of incidence. Apart from that, the proposed approach allows local excitation of spin currents if the upper layer is patterned to get a step like thickness profile. In this case the area of spin current generation is determined by the size of the pattern step and not by the laser beam diameter. Therefore, one can avoid tightly focused beams making optical resonances broader but rather deal with wide beams close to a plane wave and, at the same time, keep resolution at the micron scales. Further progress could be made if the structure surface is nanopatterned by a grating to provide surface mode excitation without a prism.

The proposed structure establishes the connection between nanophotonics and spintronics and paves a way for synergy of these two urgent branches of modern physics.

# ACKNOWLEDGEMENTS

This work is financially supported by the Russian Foundation for Basic Research (RFBR, Grant No. 18-52-80038). PVS and AKZ acknowledge support by RSF (project N 17-12-01333), YS acknowledges support by NSFC-BRICS STI Framework Program (Grant No. 51861145309), and VGA acknowledges support from DST, India (DST/IMRCD/BRICS/PilotCall2/Optonanospin/2018 (G)).

## VI. Supplementary. Materials and methods

Supplementary S1. Sample design and fabrication

The photonic crystals (PCs) were designed to enable the excitation of the optical surface mode in the vicinity of the total internal reflection angle for the glass-air interface. The parameters of the PCs were selected to achieve close to the quarter-wavelength layer thickness, and thus, to have the center of the PC bandgap at the corresponding total internal reflection angle and incident light wavelength. At the same time, the layer widths were tuned to minimize the integral thickness of the PCs [32]. This is important, since for the p-polarization one needs to deposit more than 10 pairs of layers in order to achieve suitable Bragg reflectance in the bandgap at oblique incidence.

The PCs were deposited on a fused silica substrates using magnetron sputtering and comprised dielectric non-magnetic tantalum pentoxide ($Ta_2O_5$, 119.4 nm) and silicon oxide ($SiO_2$, 164.4 nm) layers. The PC for the BIG+Pt sample had 16 pairs of layers; whereas the Py+Ta sample PC had 14 pairs of layers with the additional $Ta_2O_5$ layer of 107.5 nm width. A layer of bismuth-substituted iron garnet ($Bi_{2.1}Dy_{0.9}Fe_{3.9}Ga_{1.1}O_{12}$, 125 nm) was deposited on top of the PC for the BIG+Pt sample. After that, the PC with iron-garnet was annealed at 600°C to transform it to ferromagnetic, and this process slightly changes its permittivity. Thin layers of Pt, Py and Ta were deposited via magnetron sputtering.

Supplementary S2. Experimental setup and data processing

Measurements were performed at room temperature. For the excitation of the SPR in a thin metal film of the magnetoplasmonic heterostructure we used the Kretschmann configuration. The sample was fixed on a $SiO_2$ prism having a base angle of 42° (which is close to the angle of total internal reflection) using an index matching oil.

The structure was illuminated by a laser diode of wavelength of 805 nm and power 50 mW. The P-polarization of the laser beam was adjusted with a polarizer, and then the beam was modulated with an optical chopper operating at a frequency of 360 Hz. Modulated laser beam was focused on the sample via a micro-objective, the diameter of the focused beam was 100 µm. The angle of incidence was varied from 30 to 60 degrees through rotation of the holder on which the sample was fixed. The angular rotation accuracy was 0.05°.

The reflected light was detected by a photodiode. The sample was placed in the in-plane external magnetic field H_ext=150mT created by the electromagnet along y-axis (fig.1a). The external magnetic field **H_ext** was modulated by a sinusoidal signal at a frequency of 1.5 kHz. The resulting optically induced voltage was measured via electrodes attached to the surface of the sample using conductive glue, and amplified by a differential voltage amplifier with a gain of 1000 and an input noise level of $1nV/Hz^{0.5}$, and then digitized using a National Instruments USB-6351 data acquisition board. Then, the signal was Fourier transformed. Due to the presence of two modulating frequencies, the frequency of the modulating external magnetic field and the optical chopper operation frequency, the resulting Fourier spectra of the signal had the corresponding sum and difference frequencies at which the amplitude of the induced voltage was measured.

Supplementary S3. Numerical simulation and structure optimization

The optical properties of the opto-spintronic structures were simulated using the impedance method proposed in [32] for layered structures. The method is based on analytical recursive calculation of the so-called optical impedances $Z = \frac{E_\tau}{H_\tau}$ of the structure using the Fresnel formulas, which allow near- and far- field characteristics of the structure, the reflectance, transmittance and absorption to be obtained analytically.

The PC-based structures were optimized to obtain the narrow and deep resonances corresponding to the surface mode excitation. However, one should bear in mind that the angular width of the laser beam must typically be focused in order to achieve higher intensity. This makes it difficult to benefit from the inherent ultra-narrow surface mode resonances. The key advantage of the proposed nanostructure is that one may tune not only the position, but also the width of the resonance. Assuming the resonance curves have Lorenzian shapes [47], then

$$R(k_x) = 1 - \frac{4\Gamma_i \Gamma_{rad}}{(\beta_{mode} - k_x)^2 + (\Gamma_i + \Gamma_{rad})^2}$$

One may observe that the resonance angular width is determined by the sum of Joule $\Gamma_i$ and radiation $\Gamma_{rad}$ losses, i.e., $\Gamma_i + \Gamma_{rad}$. However, in any structure it is possible to achieve deep resonance with $R_{min} \approx 0\%$ by tuning the structure to make $\Gamma_{rad} = \Gamma_i$, since $\Gamma_{rad}$ is directly controlled by the number of layers in the PC structure.

The purple curve in Fig. 3a shows a nanostructure with wider surface optical mode resonance. The PC comprises 14 layers of 109 nm thick $Ta_2O_5$ and 181 nm thick $SiO_2$ covered with 120 nm of BIG and 1 nm of Pt. Tuning the thickness of the terminating layer results in a shift of the resonant wavelength by 1 nm per approximately 1nm change in the thickness of the considered PC.

.

Supplementary S4. PC parameters variation under annealing

PC covered with bismuth-substituted iron-garnet requires annealing at 600°C to transform BiIG to the ferromagnetic phase. The red shift of the photonic crystal bandgap edges (~50 nm approximately) was observed during this process as shown in Fig.S4.1. This effect was treated as the variation of the layer's permittivity values (2-3%) probably due to the additional oxidation process. The numerical models were corrected to match the new measured spectra. At the operating wavelength of 805 nm $\varepsilon_{BIG}=5.52+0.018i$, $\varepsilon_{Ta2O5} = 4.6200 + 0.0016i$, $\varepsilon_{SiO2} = 2.1911 + 0.0007i$.

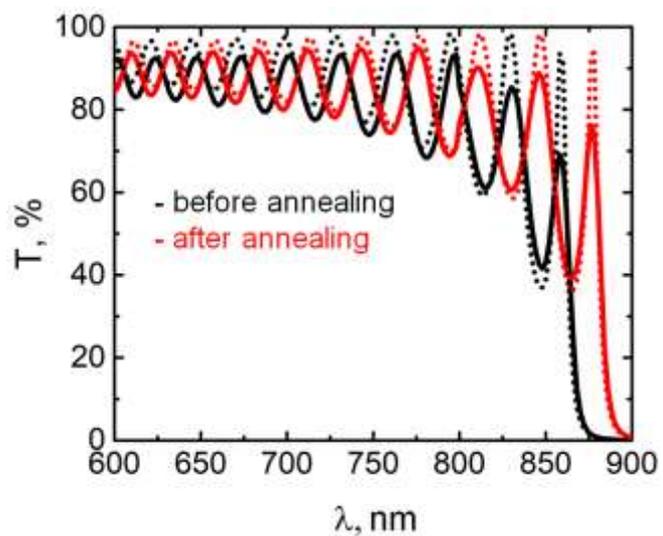

Figure S4.1 The transmittance spectra of the photonic crystal before (black color) and after (red color) annealing (600°C, 10 min) measured (solid lines) and calculated (dotted lines) at normal incidence of light.